\documentclass[journal]{IEEEtran}

\ifCLASSINFOpdf
\else
\fi
\usepackage[pdftex]{graphicx}
\usepackage{amsmath,booktabs,subfig}

\begin{document}
\title{Real-Time Betatron Tune Correction with the Precise Measurement of Magnet Current}
\author{Yoshinori~Kurimoto$^{*}$,
       Tetsushi~Shimogawa and Daichi~Naito \thanks{The authors are with the Magnet and Power Converter Group, Accelerator Laboratory, High Energy Accelerator Research Organization, Department of Accelerator Science, Graduate University for Advanced Studies (SOKENDAI),
 Tokai-mura, Ibaraki, Japan.}% <-this % stops a space
\thanks{* Y.~Kurimoto is the corresponding author. Email: kurimoto@post.j-parc.jp}}% <-this % stops a space
\maketitle

\begin{abstract}
The betatron tune, which is defined as the number of transverse oscillations in one turn of a ring accelerator, is one of the 
most important parameters. An undesired betatron tune increases the amplitude of the transverse oscillation so that many particles are lost from the ring sooner than designed. Since a betatron tune is controlled by the magnetic fields in the ring, the ripple of the magnet current directly displaces the betatron tune from its designated value. We have developed a system that corrects the betatron tune displacement using the measured magnet current at the J-PARC (Japan Proton Accelerator Research Complex) Main Ring. We adopted Field Programmable Gated Arrays (FPGA) to convert from the measured magnet current to the betatron tune in real time. Using the system, we have decreased the fluctuation of the betatron tune $\sigma_{\nu}$  from $5.2\times10^{-4}$ to  $3.3\times10^{-4}$ at the frequencies less than 250 Hz.
\end{abstract}


% \fi
\IEEEpeerreviewmaketitle

\section{Introduction}
\label{sec:intro}
\IEEEPARstart{P}articles in an accelerator generally oscillate in directions perpendicular to the beam. This transverse oscillation is 
called the betatron oscillation\cite{chao_ripple}.  The betatron oscillation is usually considered in two independent directions which are the horizontal and vertical directions. In this paper, the horizontal and vertical directions are indicated by $x$ and $y$, respectively. The transverse position of a particle in an accelerator is described as 
\begin{equation}\label{eqbeta}
x,y=\sqrt{\beta_{x,y}(s)\epsilon_{x,y}}\cos[\phi_{x,y}(s)+\phi_{0,x,y}]
\end{equation}
where $s$ is the distance along the reference orbit, $\phi_{x,y}(s)$ is the phase function, $\epsilon_{x,y}$ and $\phi_{0,x,y}$ are constants, and $\beta_{x,y}(s)$ is referred to as the betatron amplitude function. For a circular accelerator, the number of the oscillation in one turn is called the betatron tune, which is usually indicated by $\nu_{x,y}$. The betatron tune $\nu_{x,y}$ is written as 
\begin{equation}\label{eqnu}
\nu_{x,y} = \frac{1}{2\pi}\int_{C}\phi_{x,y}(s)ds 
\end{equation}
, where the $C$ indicates the line integral along the ring. The design and control of the betatron tune is one of the most important tasks for ring accelerator operation. Inappropriate betatron tunes, which satisfy the resonance condition described as 
\begin{equation}\label{eqreso}
l\nu_x + m\nu_y = n,  (l,m,n : integer)
\end{equation}
may excite amplitude growth of the transverse oscillation which causes beam losses. \\
In general accelerator tunnels, many magnets including dipole (bending), quadrupole and sextupole are located along the ring.
These magnets are in charge of controlling the parameters of the betatron oscillation. Many of these magnets are electromagnets that consists of magnetic materials and coils.  Since the magnetic field of an electrode-magnet is controlled by the current through the coil, the precise current regulator is the key to the stable operation\cite{kurimoto2014precise}. However, the magnet current regulators can not be always optimized for precise regulation since accelerator facilities are usually multi-purpose. For instance, the J-PARC Main Ring (MR)\cite{doi:10.1093/ptep/pts071}  provides proton beams to experiments via two different extraction modes. One is called ``the fact extraction'' (FX), which is used for the neutrino production. Since the neutrino production requires many primary protons, the beam is accelerated as fast as possible, and immediately extracted in the FX mode. To ramp the magnets faster, high voltage current regulators is mandatory. The other is called ``the slow extraction'' (SX), which is used for the hadron and nuclear experiments.  In these experiments, temporal uniformity of the extracted beam intensity is required so that the number of charged particles that come to their detectors can be limited within their data acquisition performance. 
In the J-PARC MR, the temporal uniformity is achieved by controlling the betatron tune to slowly approach the resonance condition. This operation is apparently related to precise current regulation. However, high-voltage current regulators generally involves large and slow semiconductor switches, which prevent precise current regulation.  In the J-PARC MR, the 4000 V 2000 A transistors are used at 450 Hz to drive the main magnets.  \\
\begin{figure}[htbp]
\begin{center}
\includegraphics[width=70mm]{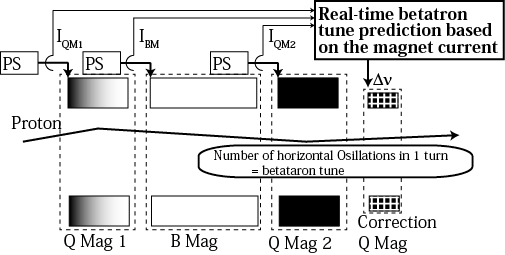}
\caption{The overview of the proposed method. ``PS'',''Q Mag'' and ``B Mag'' indicate a powex
r supply (current regulator), a quadruple magnet and a bending magnet, respectively.}
\label{fig:Concept}
\end{center}
\end{figure}
For more precise control of the betatron tune with the high voltage current regulators, we have developed a system that corrects the betatron tune displacement using the measured magnet current. The schematic overview is shown in Figure~\ref{fig:Concept}.
In the system, the betatron tune is predicted by the measured magnet current in real time. 
The predicted betatron tune is used to drive the correction quadrupole magnet. For the real-time conversion from the magnet current to the betatron tune, FPGAs are adopted. The reason why the system must have real-time feature is that the magnet current ripples are generally not predictable since the regulators are affected by the fluctuation of the main AC grid voltage and frequency.  The advantage of using the measured magnet current is non-destructive measurement for the beams. The magnet current is continuously measured for the feedback control of the regulators. Hence, the magnet current is always available as long as the regulators are under operation. On the other hand, the direct measurement of the betatron tune generally involves the excitation of transverse oscillations, which causes additional beam losses. \\
The paper is organized as follows. In Section \ref{sec:setup}, we wll describe the experimental setup of our developed system including the details of the hardware. 
In Section \ref{sec:predicted}, the betatron tune predicted by the system will be compared 
with the direct measurement.
The result of the tune correction with the system will be shown in Section \ref{sec:experiment}. Finally, we will summarize this paper in Section \ref{sec:summary}.        

\section{Experimental Setup}
\label{sec:setup}
\subsection{Overview}
\label{sec:overview}
\begin{figure}[htbp]
\begin{center}
\includegraphics[width=80mm]{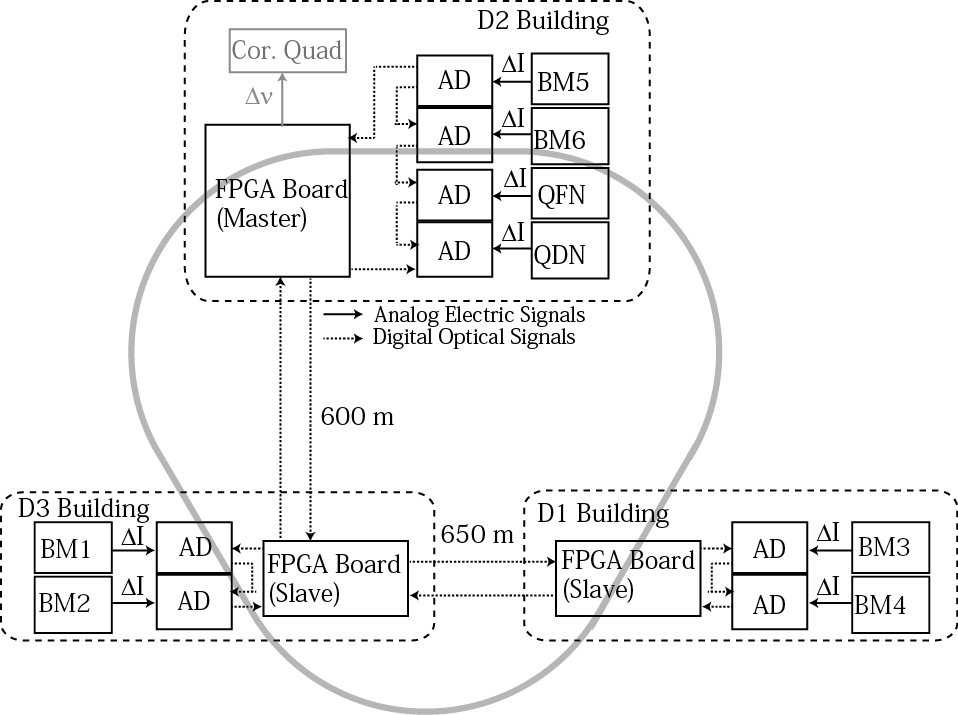}
\caption{The setup of the proposed real-time betatron tune correction system at the J-PARC MR.}
\label{fig:ripplecanceler}
\end{center}
\end{figure}
\begin{figure}[htbp]
\begin{center}
\includegraphics[width=80mm]{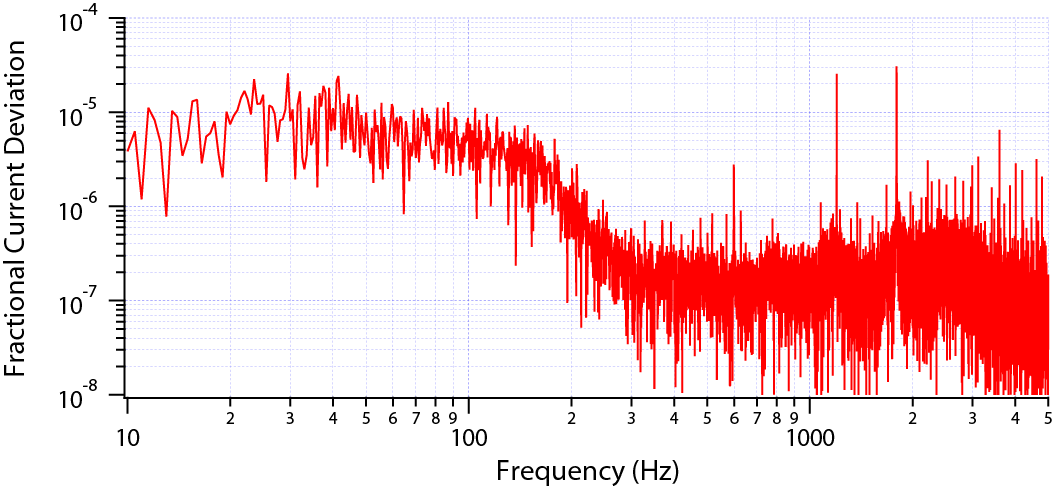}
\caption{The FFT of the current deviation of the bending magnet power supply in the J-PARC MR}
\label{fig:DIBMJPARC}
\end{center}
\end{figure}

The overview of the experimental setup is shown in Figure~\ref{fig:ripplecanceler}. The J-PARC MR uses 96 bending magnets and 216 quadrupole magnets. Six current regulators are used for all bending magnets while 11 for quadrupole magnets. All the 6 regulators for the bends and only the largest 2 regulators for the quadrupoles make dominant contribution to the betatron tune. This is the reason why we choose the 8 current regulators for the system. The current regulators of the main magnets are located in three different buildings (D1, D2 and D3), which are approximately 1 kilo-meters apart from each other. On the other hand, only one current regulator of the corrector is located at the one of them (D2). Therefore, the measured current of each regulator must be sent to the building where the corrector is located. For this purpose, we use existing optical fibers between the buildings. The fiber length is approximately 600 meters. Since there is no fiber between the D2 and D1, the signals from the D1 must go through the D3 before reaching the corrector located at the D2. The maximum delay caused by the data transfer between the buildings is approximately 6 $\mu$s, which is ignorable in comparison with the timescale of $\Delta I$. The frequency-domain $\Delta I$ of the current regulator for the bends is shown in Figure~\ref{fig:DIBMJPARC}. The dominant components are concentrated at the frequencies less than 100 Hz.

Each current regulator has analog monitor ports including the measured output current $I_{out}$ and the current deviation. The current deviation is defined as $\Delta I = I_{ref}-I_{out}$ where the $I_{ref}$ is the reference current. The $I_{out}$ and $\Delta I$ are converted to digitized optical signals using the analog-to-digital converter board (AD board) located at each current regulator. The details of the AD board will be described in Section~\ref{sec:AD board}. The FPGA board located at each building collects the $I_{out}$ and $\Delta I$ signals from each AD board, and sends them to another building. At the final FPGA board at the D2, All $\Delta I$ signals are collected and converted to the betatron tune displacement, which is sent to the corrector. This is how we correct the betatron tune using the measured magnet currents.  The hardware description of the FPGA board will be in Section~\ref{sec:fpgahard}. Apart from the $\Delta I$ for the correction, the collected $I_{out}$ and $\Delta I$ signals are sent to the upper network system via the Ethernet port  of the FPGA board. These are used as the on-line monitors of the current regulators. The way to send the collected signals to the upper system will be described in Setction~\ref{sec:fpgafirm}. 
\subsection{AD board}
\label{sec:AD board}
\begin{figure}[htbp]
  \begin{center} 
    \includegraphics[width=8.0cm]{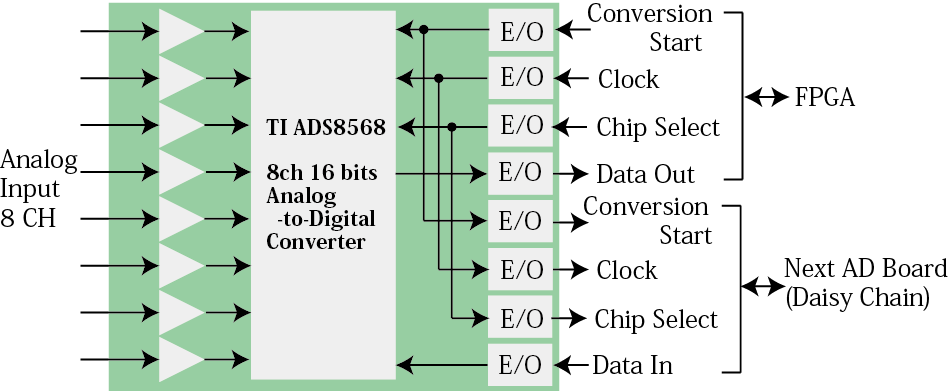}
    \caption{Concetual block diagram of the developed AD board.} 
    \label{fig:ConceptADBoard} 
  \end{center}
\end{figure}

\begin{figure}[htbp]
  \begin{center} 
    \includegraphics[width=7.0cm]{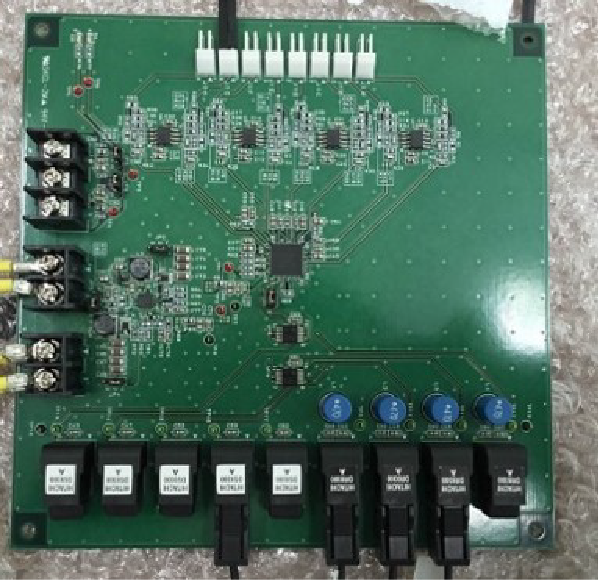}
    \caption{Picture of the AD board.} 
    \label{fig:PictureADBoard} 
  \end{center}
\end{figure}

The conceptual block diagram and picture of the AD board are shown in Figure~\ref{fig:ConceptADBoard} and Figure~\ref{fig:PictureADBoard}, respectively. 
The 8-channels 16-bit analog-to-digital converter (Texus Instruments ADS8568) is adopted
for the AD converter IC. The sampling frequency is set to 10 kHz in the system.
%The board was originally developed for the controllers of power converters.
%The power control generally requires many feedback parameters, for example three phase voltages and currents. This is the reason %of the 8 channel analog inputs despite of the only two ($\Delta I$ and $I_{out}$) for this system. 
The digital inputs for the control and digitized data output are all optical. 
Although the AD converter IC itself provides both serial and parallel digital outputs, the board only provides the serial digital output. The multiple boards can forms into ``Daisy Chain'' so that an additional board can be easily included in the existing system by connecting the new board to the tail of the system. The upper four optical ports in Figure~\ref{fig:ConceptADBoard} are for the communication to the master. The AD conversion is triggered by the ``Conversion Start'' signal. It takes approximately 1.7 $\mu$s until the completion. After the AD conversion, the digitized data are transfered to the master via the ``'Data Out' port when the ``Chip Select'' port is asserted. The data transfer starts with the most significant bit of the first channel. The data is updated to the next bit in synchronization with the 20 MHz clock on the ``Clock'' port. Not only the data from the other channels but also from the other slave boards are also transfered through the ``Data Out'' port. The lower four optical ports in the figure are for the communication to the slave AD boards.  The ``Conversion Start'', ``Clock'' and ``Chip Seclect'' signals are sent to the slave boards 
using the first three ports out of the lower four. The digitized data of the slave boards are received via the ``Data In'' port.

\subsection{Hardware Design of the FPGA board}
\label{sec:fpgahard}
\begin{figure}[htbp]
  \begin{center} 
    \includegraphics[width=8.0cm]{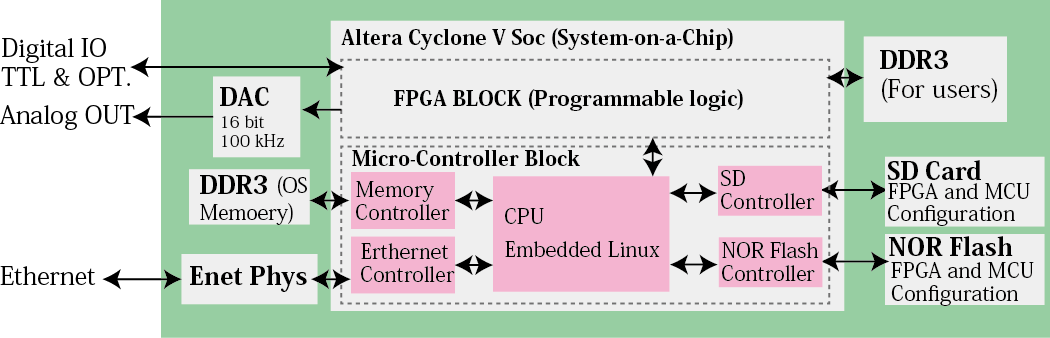}
    \caption{Conceptual block diagram of the developed FPGA board.} 
    \label{fig:ConceptBoard} 
  \end{center}
\end{figure}
\begin{figure}[htbp]
  \begin{center} 
    \includegraphics[width=7.0cm]{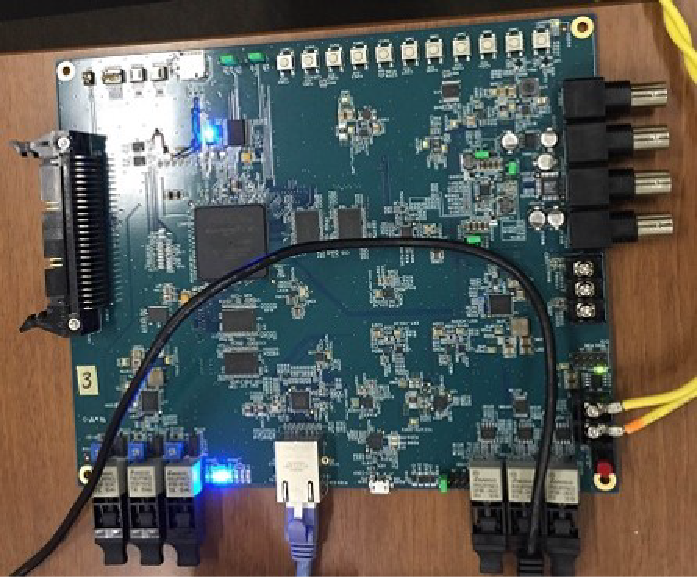}
    \caption{Picture of the developed board.} 
    \label{fig:PictureBoard} 
  \end{center}
\end{figure}
The conceptual block diagram and picture of the developed board are shown in Figure~\ref{fig:ConceptBoard} and Figure~\ref{fig:PictureBoard}, respectively. We have adopted a System-on-Chip (SoC) FPGA for the intelligent component of the board. 
In a single SoC FPGA chip, a processor, memory controllers and peripherals are integrated with the FPGA block. Hereinafter, these are collectively referred as to ``micro-controller block'' against the FPGA block. The details of the chosen SoC FPGA 
will be described in Section~\ref{sec:SOCFPGA}. General purpose digital IO ports and 4-channel analog outputs are available 
in the board. The specific configuration of the SoC FPGA is loaded from a SD card or NOR flash memory. A single gigabit ethernet port is also available so that we can communicate with the board via a network. This FPGA board was originally developed as the multi-purpose intelligent I/O board for the J-PARC MR. The other applications are described in \cite{SoC}.
\paragraph{SoC FPGA} \label{sec:SOCFPGA}
We have chosen a device from Altera Cyclone V SX Soc families. The device name is 5CSXC6.
The FPGA block includes 41509 active logic modules, 5570 Kb internal memory, 112 variable-precision
DSP blocks, 288 user I/Os. The micro-controller block includes dual processor cores (ARM Cortex-A9 MPCore),
 a hardened memory controller. 
The FPGA block is for functions specific to accelerator components. This block is very suitable especially for real-time applications such as power supply control, beam position feedback, low-level RF control and so on. This is because the functions in the FPGA block are configured as hardened circuits and the timing properties such as throughput and latency are exactly determined. An Linux operating system (OS) is executed in the micro-controller, which can internally communicate with the FPGA block. The DDR (Double Data Rate) 3 memory devices attached to the micro-controller block are used for the OS expansion. We execute one user program in the OS. The program can start and stop the operation of the functions implemented in FPGA block. In addition, it plays a role in changing the parameters of the functions. This gives us flexibility for designing circuits in the FPGA block with variable parameters and states. The program also monitors the parameters and states. Some of the parameters and states are required for other accelerator components and must be remotely changed. To carry out these tasks, the program also works as a network socket server so that we can access the board from the outside through an Ethernet port.
\paragraph{User Memory Device}
We have adopted 1GB DDR3 as a user memory device. This large memory storage is very useful especially for accelerators with relatively longer repetition periods. In those accelerators, the states of many accelerator components must vary with the proceeding of beam acceleration. For example, we must vary the currents for main magnets corresponding to the particle momentum. Therefore, we need to store not single value but a table of the reference current values. In additions, we usually need to apply some corrections to the accelerator components and the correction values may also be tables. 
\subsection{Firmware Design of the FPGA}
\label{sec:fpgafirm}
\begin{figure*}[htbp]
  \begin{center} 
    \includegraphics[width=16.0cm]{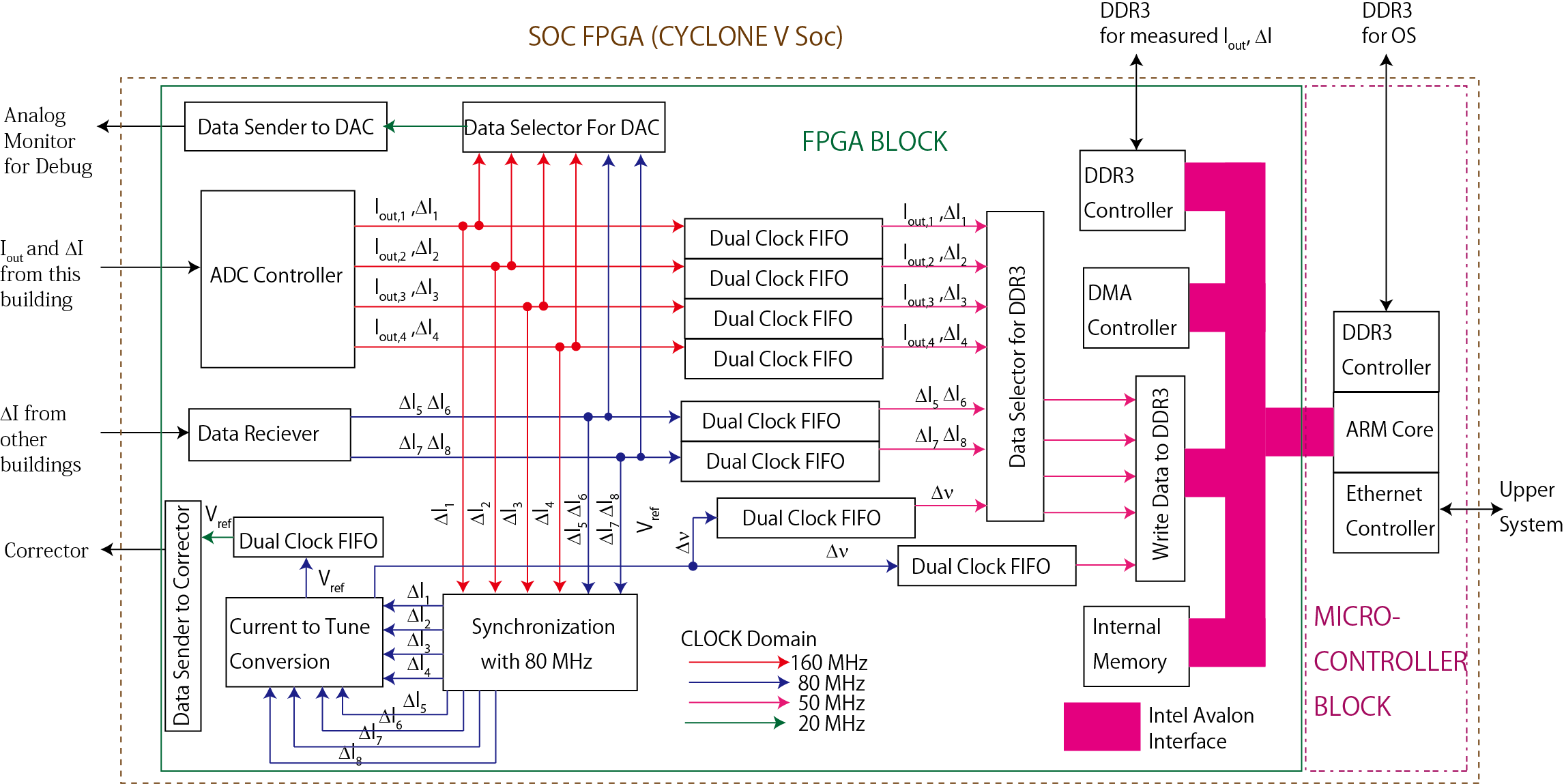}
    \caption{Conceptual block diagram of the developed firmware on the SoC FPGA.} 
    \label{fig:Firmware} 
  \end{center}
\end{figure*}
Figure~\ref{fig:Firmware} shows the block diagram of the firmware on the SoC FPGA.
This firmware corresponds to the final FPGA board located at the D2 building. 
There are two main data ($I_{out}$, $\Delta I$) streams in the firmware. One is for the conversion from the $\Delta I$ signals to the betatron tune displacement.  
This corresponds to the left part of Figure~\ref{fig:Firmware}.  The betatron tune 
displacement can be calculated as the linear combination of the $\Delta I$ signals ($\Delta \nu = \sum^{8}_{i=1} \alpha_{i}\Delta I_{i}$). The coefficients $\alpha_{i}$ 
, which are determined by the optics models of the J-PARC MR, are stored in the internal memory. After the conversion, the calculated tune displacement is sent to the corrector.
The other stream is for the on-line monitoring. This is described in the right part of the figure. The received $\Delta I$ and $I_{out}$ are temporally stored in the DDR3 connected to the FPGA block. Once the data for one accelerator cycle (2.48 - 5.2 sec) are stored, 
they are transfered into the DDR3 connected to the micro-controller block with the DMA (Direct Memory Access) control. The network program running in the micro-controller block sends the transfered data to the upper system via the Ethernet port. The communication between the FPGA and micro-controller blocks can be done via a standard interface which is called ``Intel Avalon Interface''.

\section{Comparison with the direct measurement}
\label{sec:predicted} 
\begin{figure}[htbp]
  \begin{center} 
   \centering
    \subfloat[The direct measurement of the betatron tune.]{\includegraphics[width=9.0cm]{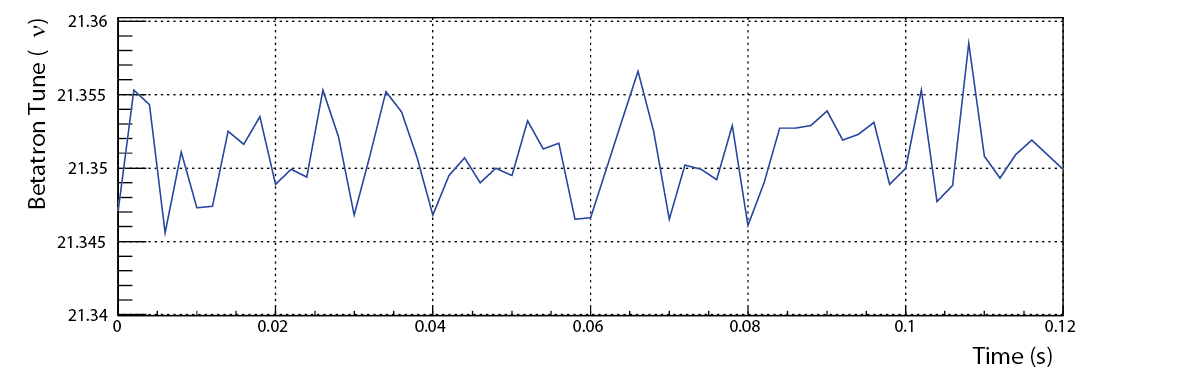}\label{fig:Measured}} 
   \\   
   \subfloat[The betatron tune prediction, which is calculated with the system (in real time)]{\includegraphics[width=9.0cm]{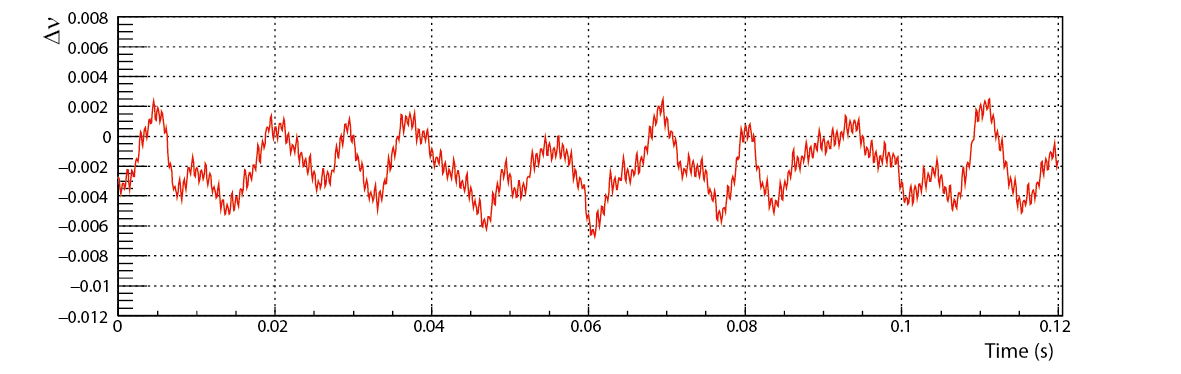}\label{fig:Predicted}}
    \caption{The comparison between the direct measurement and the prediction} 
    \label{fig:Compare} 
    %\caption{The betatron tune prediction, which is calculated with the system (in real time)} 
    %\label{fig:Predicted} 
  \end{center}
\end{figure}

For the direct measurement of the betatron tune, we measure the frequency  
of the transverse oscillation excited by the stripline kickers\cite{Toyama:2010tma}. This measurement is destructive and can not be done for the user operation. Therefore, it is important to compare the predicted tune with the direct measurement during the beam study time.  Figure~\ref{fig:Compare} shows the comparison between the direct measurement and the prediction. Figure~\ref{fig:Measured} shows the direct measurement while Figure~\ref{fig:Predicted} shows the prediction using the system. The corrector was turned off for this comparison. The tendencies of the time variation are roughly same although there are some discrepancies between them. There are possible reasons for the discrepancies. One possibility is the precision of the direct measurement. The transverse excitation at some time windows is not enough so that the betatron sideband can be hardly found. Another reason is that the prediction is not done by the all magnet current regulators as mentioned in Section~\ref{sec:overview}.    

\section{Betatron Tune Correction}
\label{sec:experiment}
\begin{figure*}[htbp]
  \begin{center} 
    \includegraphics[width=16.0cm]{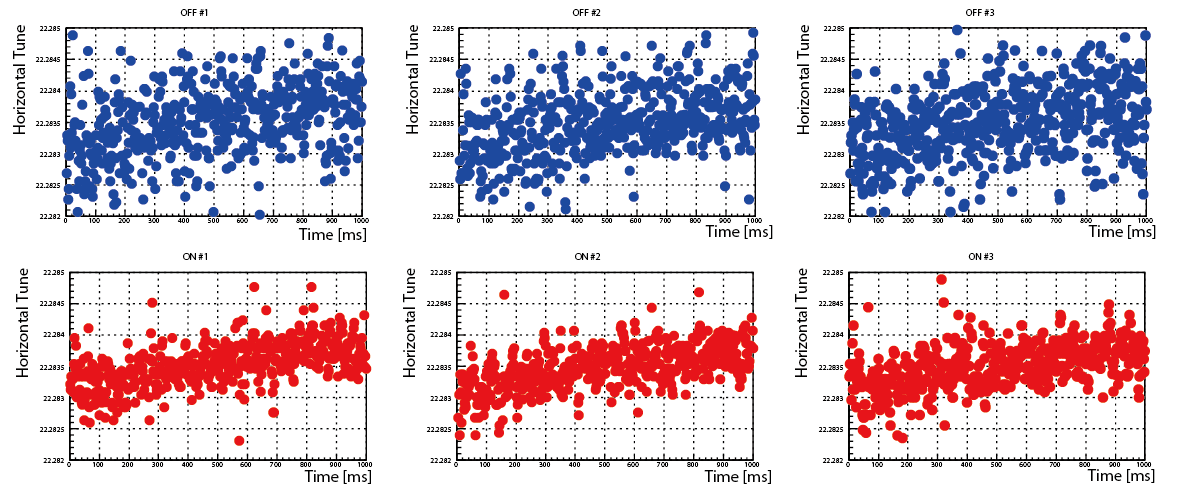}
    \caption{The measured betatron tunes as a function of time. The upper three plots correspond to the measurements without the system while the lower three with the system.} 
    \label{fig:Graph} 
  \end{center}
\end{figure*}

\begin{figure*}[htbp]
  \begin{center} 
    \includegraphics[width=16.0cm]{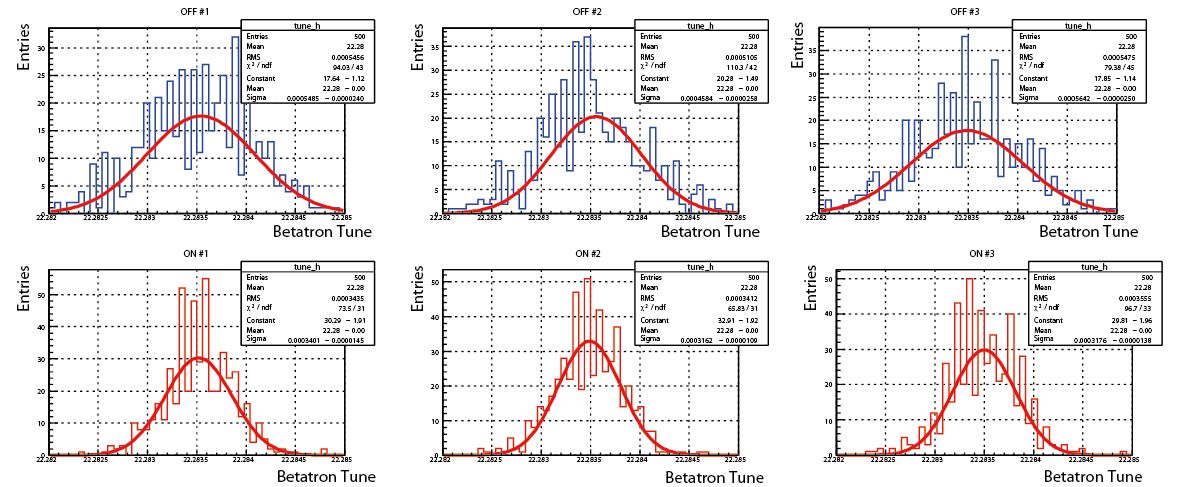}
    \caption{The histograms of the measured betatron tunes. The upper three plots correspond to the measurements without the system while the lower three with the system.} 
    \label{fig:Hist} 
  \end{center}
\end{figure*}

To examine the effect of the system, we measured the betatron tune three times in the SX mode with the corrector turned on. 
%\footnote{As we mentioned in Section \ref{sec:intro}, this ``direct'' measurement is destructive for the beams. Therefore, this can not be done during the user operation}.  
For the comparison, the betatron tune without the corrector turned off is also measured. The upper three plots 
in Figure~\ref{fig:Graph} show the measured betatron tunes with the corrector turned off while the lower three with turned on. The interval between the measured points is 2 ms. Hence, these plots are sensitive to the frequencies below 250 Hz which cover the dominant frequencies of the $\Delta I$ ($\leq$ 100 Hz).  
In the actual user operation, the beams are controlled to the resonance condition and slowly extracted during these measurement periods. 
Therefore, the fluctuation of the measured betatron tune in these plots must be smaller with the system. To check this more clearly, we also show the projection of these plots (histograms) in Figure~\ref{fig:Hist} and performed the Gaussian fitting to these histograms. The upper three histograms correspond to the measurements with the corrector turned off while the lower three with turned off. The standard deviations of the fitted Gaussian are shown in Table~\ref{table:sigma}. The system reduces the fluctuation by approximately 43 \%  
\begin{table}[hbtp]
  \caption{The fluctuation of the betatron tune.}
  \label{table:sigma}
  \centering
  \begin{tabular}{ccccc}
    \hline
     & Shot 1  &  Shot 2 & Shot 3 & Average \\
    \hline \hline
    Corrector Off  &  $5.5\times10^{-4}$ &$4.6\times10^{-4}$  & $5.6\times10^{-4}$& $5.2\times10^{-4}$ \\
    Corrector ON  &  $3.4\times10^{-4}$  &  $3.2\times10^{-4}$ &  $3.2\times10^{-4}$ & $3.3\times10^{-4}$ \\
    \hline
  \end{tabular}
\end{table}
\section{Conclusion}
\label{sec:summary}
The betatron tune, which is defined as the number of transverse oscillations in one turn of a ring accelerator, is one of the most important parameters. An undesired betatron tune increases the amplitude of the transverse oscillation so that many particles are lost from the ring sooner than designed. Since a betatron tune is controlled by the magnetic fields in the ring, the ripple of the magnet current directly displaces the betatron tune from its designated value. \\
We have developed a system that corrects the betatron tune displacement using the measured magnet current at the J-PARC MR. We adopted Field Programmable Gated Arrays (FPGA) to convert from the measured magnet current to the betatron tune in real time. The $\Delta I$ signals from the 8 magnet current regulators are used to calculate the betatron tune displacement. We have only one corrector magnet while these 8 regulators are located in different three buildings. Hence, the digitized $\Delta I$ signals at each regulator are sent to the building where the corrector is located through the optical fibers. \\   
We performed the direct measurement of the betatron tune with the system. The measurement shows
that the system can decrease the fluctuation of the betatron tune $\sigma_{\nu}$ from $5.2\times10^{-4}$ to $3.3\times10^{-4}$ at the frequencies less than 250 Hz.

\section*{Acknowledgment}
We thank to all members of the power converter group and the slow extraction group in the J-PARC MR for their continuous encouragement to this work. We are 
grateful especially to Mr. Sagawa's assistance for the board development and installation.
 
% Can use something like this put references on a page
\ifCLASSOPTIONcaptionsoff
  \newpage
\fi
\bibliographystyle{IEEEtran}
\bibliography{./IEEEabrv,./myfile}

\end{document}